\documentclass[doublecol,amsmath]{epl2}
\usepackage{graphicx}    
\usepackage{hyperref}    
\usepackage{amsmath}
\title{Phosphorene nanoribbons}

\author{A. Carvalho\inst{1} A. S. Rodin\and \inst{1} A. H. Castro Neto\and \inst{1,2}}

\institute{

\inst{1}Graphene Research Centre and Department of Physics, National University of Singapore, 117542, Singapore\\

\inst{2}Boston University, 590 Commonwealth Ave., Boston MA 02215\\
}

\abstract{
Edge-induced gap states in finite phosphorene layers are examined using analytical models and density functional theory.
The nature of such gap states depends on the direction of the cut.
Armchair nanoribbons are insulating, 
whereas nanoribbons cut in the perpendicular direction (with zigzag and cliff-type edges)
are metallic, unless they undergo a reconstruction or distortion with cell doubling, 
which opens a gap.
All stable nanoribbons with unsaturated edges have gap states that can be removed by hydrogen passivation.
Armchair nanoribbon edge states decay exponentially with the distance to the edge
and can be described by a nearly-free electron model.
}

\begin{document}
\maketitle

Black phosphorus is a van der Waals-bonded layered material, 
and the most stable form of of this element.
Its peculiar orthorhombic structure resembling corrugated cardboard sets  it apart 
from other two dimensional materials such as graphene and MoS$_2$.
Unlike those, it is highly anisotropic with respect to the directions in-plane,\cite{rodin,liu,fei,low}
and therefore it offers new terrain to the exploration of physical phenomena
in two-dimensional materials.

The isolation of the monolayer form, named phosphorene, 
has been recently achieved by mechanical cleavage followed by plasma thinning\cite{qiao}
and multi-layered material 
is now routinely produced by exfoliation.\cite{li,koenig}
Both have been theoretically predicted to be  direct-gap or nearly direct gap semiconductors
with energy gaps ranging from 2~eV to 0.8~eV.\cite{tran}
However, the bandstructure is far from typical.
The effective masses for carriers are at least one order of magnitude higher along the direction of
the to the zigzag ridges ($y$) than along the perpendicular direction ($x$).
And the electron and hole masses are nearly the same along $x$,
but very asymmetric along $y$.
Further, due to the weak interaction between second and third nearest neighbors,
it can easily be deformed under uniaxial compression,
becoming, in sequence, an indirect gap semiconductor, a semimetal, a metal, and ultimately undergoing a transition to a square lattice.\cite{rodin, liu,peng}

Phosphorene can, in principle, be cut and tailored into derived nanostructures\cite{guoNR,pengNR,maityNR}.
An interesting feature of finite systems is the possibility of edge states. 
These states are localized at the system-vacuum interface and decay exponentially away from it. 
The nature of the edge states depends not only on the crystal structure of the system, but also on the way it terminates. 
Understanding the physics of nanoribbon edges is fundamental 
for predicting the behavior of real finite systems
and for designing more complex nanostructures, such as
nanotubes and voids.
In this paper, we analyze the electronic states induced by finite edges.
We show that armchair edges, cut along the $x$ direction, invariably lead to localized edge states, decaying exponentially towards the bulk.
In contrast, edges cut along the $y$ direction were found to introduce less localized states for nanoribbons.
All the nanoribons are semiconducting in the large width limit.

We first discuss the origin of these localized gap states.
In our previous work~\cite{rodin}, we have shown that tight-binding has a limited utility for black phosphorus. Therefore, we will perform our analysis here using the nearly-free electron model (NFEM). Before proceeding with the particular phosphorene case, we provide a derivation of the edge states for the reader's convenience following a standard procedure.

We consider edge states in the gap of the infinite monolayer. Since these states are exponentially decaying, they must have an imaginary momentum component perpendicular to the edge. Assuming that the system is located in the negative half-plane, the wavefunction inside it is given by
\begin{equation}
\Psi^I = e^{iq u+ i(-ip)w}\sum_n A_n e^{i nL_w w}\,,
\label{eqn:Psi_I}
\end{equation}
where $u$ is the coordinate along the edge and $w$ is perpendicular. $L_w=2\pi/a_w$ is the smallest reciprocal vector transverse to the edge so that the summation runs over all the reciprocal vectors along $\hat w$, in accordance with the Bloch theorem. To keep the notation as clean as possible, we introduce 
\begin{equation}
E_n^q = \frac{\hbar^2\left[q^2+\left(nL_w-ip\right)^2\right]}{2m} = (E^q_{-n})^*\,.
\end{equation}

Consider a gap at the $\Gamma$ point between $2N$th and ($2N+1$)th bands. This means that the states in the vicinity of the gap are primarily defined by $N$th and $-N$th harmonics. Writing down the Bloch equation for only these two harmonics gives
\begin{equation}
H_N\Psi_N = [E-V_0]\Psi_N\,,\quad H_N = \begin{pmatrix}
E^q_{-N}&V_{2N}
\\
V_{2N}&E^q_N
\end{pmatrix}\,.
\label{eqn:HM}
\end{equation}
In the equation above, $V_n = V_{-n}= V_{\mathbf G = nL_w\hat w}$, where $V_\mathbf{G}$ is the Fourier component of the lattice potential. Diagonalizing the Hamiltonian in Eq.~\eqref{eqn:HM} yields
\begin{equation}
\frac{E}{F} = N^2L_w^2+p^2+q^2+\frac{V_0}{F}\pm\sqrt{\frac{V_{2N}^2}{F^2}-4N^2L_w^2p^2}\,,
\label{eqn:E}
\end{equation}
where $F = \hbar^2/(2m)$. The eigenvectors are
\begin{equation}
\Psi_N = \begin{pmatrix}
e^{i\delta_N}\\e^{-i\delta_N}
\end{pmatrix}\,,\quad \delta_N = \frac{1}{2}\sin^{-1}\left(-\frac{2pNL_wF}{V_{2 N}}\right)\,.
\label{eqn:Delta}
\end{equation}
with
\begin{equation}
2pNL_w\tan\delta_N = -\frac{V_{2N}}{F}\pm\sqrt{\frac{V_{2N}^2}{F^2} - 4N^2L_w^2p^2}\,.
\label{eqn:tan_delta}
\end{equation}
From this, we get the following form for the internal wavefunction:
\begin{equation}
\Psi_N^I \sim e^{i q u+p_Nw}\cos\left(N L_w w+\delta_N\right)\,.
\end{equation}

By setting the outside wavefunction to
\begin{equation}
\Psi^O = e^{iqu-\sqrt{q^2-E/F}w}
\end{equation}
one can show that the momentum $p_N$ is determined from
\begin{equation}
 p_N-NL_w\tan(NL_wW+\delta_N)=-\sqrt{q^2-\frac{E_N}{F}}\,,
\label{eqn:Solution}
\end{equation}
where $W$ is the coordinate of the boundary. Defining $\alpha = NL_wW$, $k = p_N/(NL_w)$, $U = V_{2N}/(FN^2L_w^2)$, and $B = V_{0}/(FN^2L_w^2)<0$ allows us to combine Eqs.~\eqref{eqn:E}, \eqref{eqn:tan_delta}, and \eqref{eqn:Solution} to get
\begin{widetext}
\begin{equation}
k-\left[\frac{\tan\alpha-\frac{2k/U}{\sqrt{1-4\frac{k^2}{U^2}}+1}}{1+\frac{2k/U\tan\alpha}{\sqrt{1-4\frac{k^2}{U^2}}+1}}\right]+\sqrt{-B-U-k^2-1+k\frac{4k/U}{\sqrt{1-4\frac{k^2}{U^2}}+1}}=0\,.
\label{eqn:Exist}
\end{equation}
\end{widetext}
Realistically $|B|\gg |U|$. In addition, $\tan\alpha$ is $\pi$--periodic so that $\alpha$ enters as 
\begin{equation}
\alpha\rightarrow (\alpha\text{ mod }\pi) - \pi\,\Theta[(\alpha\text{ mod }\pi)- \pi/2]\,.
\label{eqn:alpha_RE}
\end{equation}
It is instructive to determine the values of $\alpha$ that allow positive real-$k$ solutions of Eq.~\eqref{eqn:Exist} for specified $B$ and $U$. This means that we need to find all values of $\alpha$ where the left-hand side of Eq.~\eqref{eqn:Exist} changes sign for $0\leq k\leq|U|/2$. To do so, we first find the values of $\alpha$ which have solutions at $k = 0$ and $k = |U|/2$. For the first case, one gets
\begin{equation}
\alpha_1 = \tan^{-1}\left[\sqrt{-1-B-U}\right]>0\,.
\end{equation}
The second case yields
\begin{equation}
\alpha_2 = \tan^{-1}\left[-\frac{2+U+\sqrt{-4-4B-U^2}\text{sign}[U]}{(-2+U)\text{sign}[U]+\sqrt{-4-4B-U^2}}\right]\,.
\end{equation}
Assuming that $|B|$ is large enough so that all the square roots are real, it is possible to show by plotting that the $\alpha$ that results in positive real $k$ is
\begin{align}
\alpha_2\leq\alpha\leq\alpha_1\,,\quad \text{if }U<0\,,
\label{eqn:U_less}
\\
-\frac{\pi}{2}\leq\alpha\leq\alpha_2\text{ or }\alpha_1\leq\alpha\leq\frac{\pi}{2}\,,\quad\text{if } U>0\,.
\label{eqn:U_greater}
\end{align}
This expression will be used to determine whether one should expect edge states in phosphorene.

In order to establish the existence of edge states, we need to find the value of $U$ and, therefore, $V_{2N}$. As there are four atoms per unit cell in phosphorene, the Fourier component $V_{\mathbf G}$ of the full potential is given by the product of the transform of a single atom potential $V_\mathbf G^A$ multiplied by the geometrical structure factor $Q_\mathbf G$. Since the atomic potential is attractive, $V_\mathbf G^A<0$.

\begin{figure}
\includegraphics{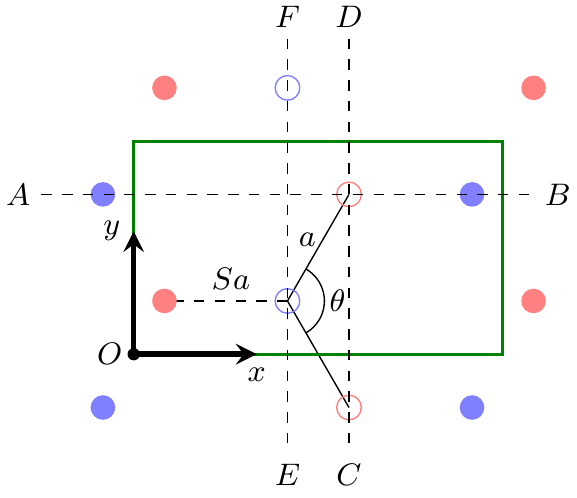}
\caption{Black phosphorus lattice with a rectangular unit cell. The dashed lines describe the three edge types: AB is for armchair; CD is for zigzag; EF is for cliff. Note that for armchair, the system is located below the cut; for zigzag and cliff, it is to the left of the cut.}
\label{fig:Lattice}
\end{figure}

The atoms inside the unit cells are located at $\pm \mathbf l_1$ and $\pm \mathbf l_2$, where
\begin{align}
&\mathbf l_1=\left(\frac{a}{2}\cos\frac{\theta}{2},\frac{a}{2}\sin\frac{\theta}{2}\right)\,,
\\
&\mathbf l_2=\left(\frac{a}{2}\cos\frac{\theta}{2}+ Sa,\frac{a}{2}\sin\frac{\theta}{2}\right) \,,
\end{align}
see Fig.~\ref{fig:Lattice} for details. The corresponding reciprocal vectors are
\begin{equation}
\mathbf G =\frac{\pi}{a}\left[\frac{n_x}{S+\cos\frac{\theta}{2}}\hat x+\frac{n_y}{\sin\frac{\theta}{2}}\hat y\right]\,,
\end{equation}
where $n_x$ and $n_y$ are integers.  If $\mathbf G$ points along one of the principal axes, so that either $n_x$ or $n_y$ = 0, one gets the following. For $n_x = 0$,
\begin{equation}
Q_\mathbf G = 4\cos\left(\frac{n_y\pi}{2}\right)=\begin{cases}
4&\text{if } n_y=4,8,12\dots
\\
0&\text{if } n_y \text{ is odd}
\\
-4&\text{if } n_y=2,6,10\dots
\end{cases}\,.
\label{eqn:Qy}
\end{equation}
On the other hand, if $n_y = 0$:
\begin{align}
Q_\mathbf G&=2\left[\cos\left(\frac{n_x\pi}{2}\left[\frac{\cos\frac{\theta}{2}}{S+\cos\frac{\theta}{2}}\right]\right)+\right.
\nonumber
\\
&\left.\cos\left(\frac{n_x\pi}{2}\left[\frac{S}{S+\cos\frac{\theta}{2}}+1\right]\right)\right]\,.
\end{align}

We are now in position to make certain predictions concerning the edge states, starting with the armchair geometry. 
The armchair edges cut phosphorene along a line parallel to the $x$ axis,
in such a way that the projection into the $xOy$ plane resembles a graphene  armchair edge. Thus, we have in this case $\hat w = \hat y$ and $L_w W =\pi/2$. In addition, we know that $N = 5$ since the gap is between the tenth and the eleventh bands. This means that the geometrical structure factor $Q = -4$, in accordance with Eq.~\eqref{eqn:Qy}. Therefore, $U\propto V_\mathbf{G} = V_\mathbf{G}^A\times Q>0$ since the Fourier component of the attractive atomic component $V_\mathbf{G}^A$ is negative. Finally, $\alpha = NL_wW = N\pi/2\rightarrow \pi/2$ for odd $N$, per Eq.~\eqref{eqn:alpha_RE}. From Eq.~\eqref{eqn:U_greater}, we conclude that odd-$N$ harmonics have real-$k$ solutions and armchair edges contain edge states.

Next, we address the zigzag and cliff edges.  Here, the cuts are made along the $y$ axis and $L_w =\pi/\left(Sa+a\cos\frac{\theta}{2}\right)$. Following our earlier assumption that the system extends to negative infinity, the boundaries of zigzag and cliff edges are at $W =\pm\frac{a}{2}\cos\frac{\theta}{2}$, respectively. This yields $\alpha = \pm N\pi\cos(\theta/2)/[2(S+\cos(\theta/2))]$. Unlike the armchair case, here the reduced $\alpha$ is between $0$ and $\pi/2$. Therefore, one needs to know the explicit form of the potential to determine $\alpha_1$ and $\alpha_2$ to see whether Eqs.~\eqref{eqn:U_less}-\eqref{eqn:U_greater} hold. Without this information, one cannot make a definite prediction concerning the existence of edge states.


We confirm this by modeling numerically the electronic band structure of 
infinitely long black-phosphorus nanoribbons.
\cite{footnote}
The nanoribbons were modeled within the framework of density-functional theory,\cite{dft} as implemented in the {\sc siesta} package.\cite{siesta1,siesta2}
The generalized gradient approximation of Perdew, Burke and Ernzerhof is used for the exchange-correlation functional.\cite{PBE}
The electronic core is accounted for by using {\it ab-initio\/} norm-conserving pseudopotentials with the Troullier-Martins parametrisation\cite{TM} in the Kleinman-Bylander form.\cite{KB}
The charge density was assumed to be independent on spin. 

The basis sets for the Kohn-Sham states are linear combinations of numerical atomic orbitals.\cite{siestabasis1,siestabasis2} 
These are a double zeta polarized basis set for P and a single zeta polarized
basis set for H.
The charge density is projected on a real-space grid with an equivalent cutoff energy of 250~Ry to calculate the exchange-correlation and Hartree potentials.
A Monkhorst-Pack~\cite{MP} scheme with at least 10 points
is used to sample the Brillouin Zone. 
The structural parameters of monolayer black phosphorus are given in Table~1.

\begin{figure*}
\includegraphics[scale=1]{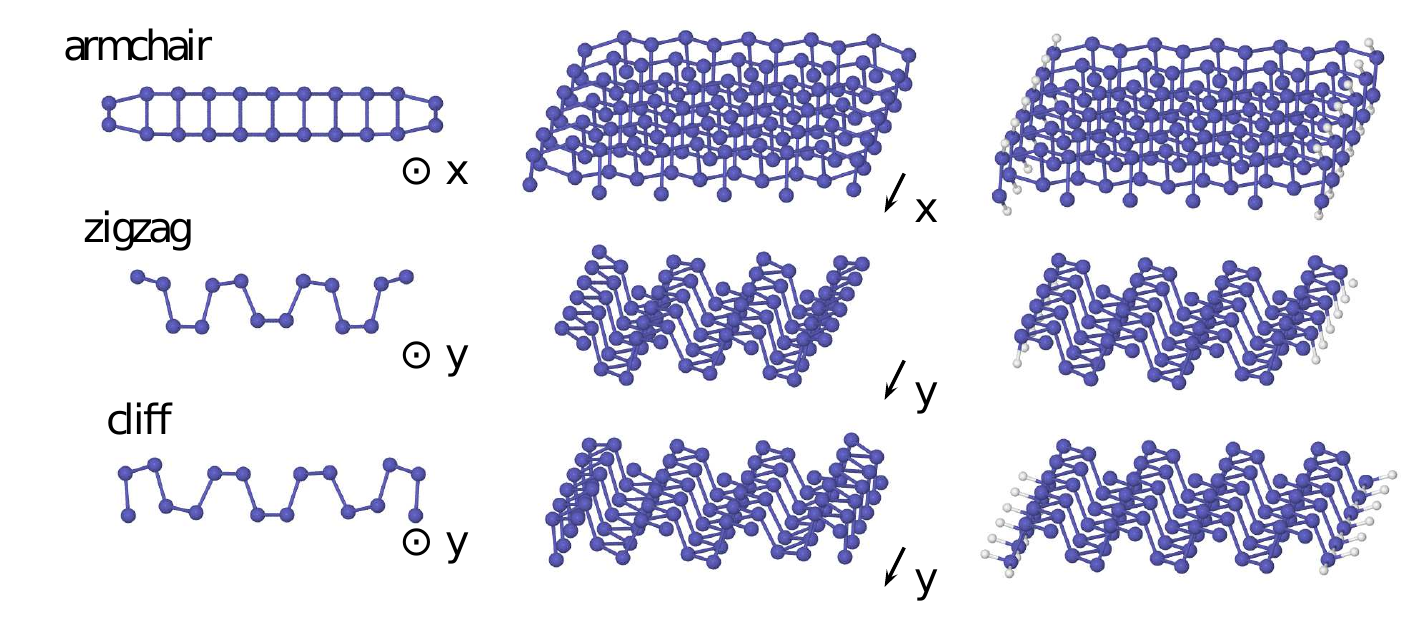}
\caption{(Color online)
Phosphorene nanoribbons (unreconstructed) with 
armchair, zigzag and cliff-edges (with $n$=11, 14 and 16, respectively).
For each type of edge, it is show, from left to right:
cross section of the unsaturated nanoribbon,
where the infinite direction runs perpendicular to the drawing; 
and perspective view of the unsaturated nanoribbon, and of hydrogen-saturated nanoribbon.
Phosphorus and hydrogen atoms are represented by large (blue) and small (white)
spheres, respectively.
\label{fig:NRs}
}
\end{figure*}

The armchair nanoribbons are  constructed 
by cutting a phosphorene monolayer parallel to the $x$ axis,
in such a way that they are symmetric with respect to the $xOz$ plane.
As they can also be seen as being composed of $n$ atomic lines, each one parallel to the $x$ axis, we use $n$ to label the nanoribbon width (Fig.~\ref{fig:NRs}).

The zigzag nanoribbons are  constructed by cutting a phosphorene monolayer parallel to the $y$ axis,
in such a way that they are symmetric with respect to the $yOz$ plane.
We also label those nanoribbons by the number ($n$) of atomic lines parallel to the $y$ axis,
where $n$ is an even number not multiple of 4 (Fig.~\ref{fig:NRs}).
Zigzag nanoribbons undergo little structural relaxation with exception of the outermost edge atoms.
However, they are sensitive to Peierls distortion\cite{maityNR} with unit cell doubling,
 which results in a modest energy gain ($<40$~meV).

Another type of edge along $y$, which we name ``cliff"-edge, is found by cutting the layer along the line [EF] in Fig.~\ref{fig:Lattice}.
As the cliff edge leaves the edge P atoms with coordination one, it leads to considerable reconstruction.
We found the reconstruction with unit cell doubling proposed in Ref.~\cite{maityNR} to be the most stable,
leading to an energy gain of 0.37 eV/unit cell per edge relative to the reconstruction respecting the
original periodicity of the nanoribbon along the $y$ direction.
Since the resulting local edge structure is very distinct from the pristine black phosphorus,
this type of nanoribbon cannot be described using our previous analytical arguments.
For cliff-edge nanoribbons $n$ is a multiple of 4.

\begin{table}
\begin{center}
\begin{tabular}{ll}
\hline
a   ($\AA$) &  2.2578  \\
Sa  ($\AA$) & 0.8014   \\
h   ($\AA$) & 2.15   \\
\hline
$\theta$ ($^\circ$) & 95.7  \\
\end{tabular}
\end{center}
\caption{Structural parameters of monolayer phosphorene:
$a$ is the length of the in-plane bond,
$Sa$ is the length of the projection of the out-of-plane bond in the plane,
and $h$ is the layer thickness.
}
\end{table}

\begin{figure}
\includegraphics[width = 8cm]{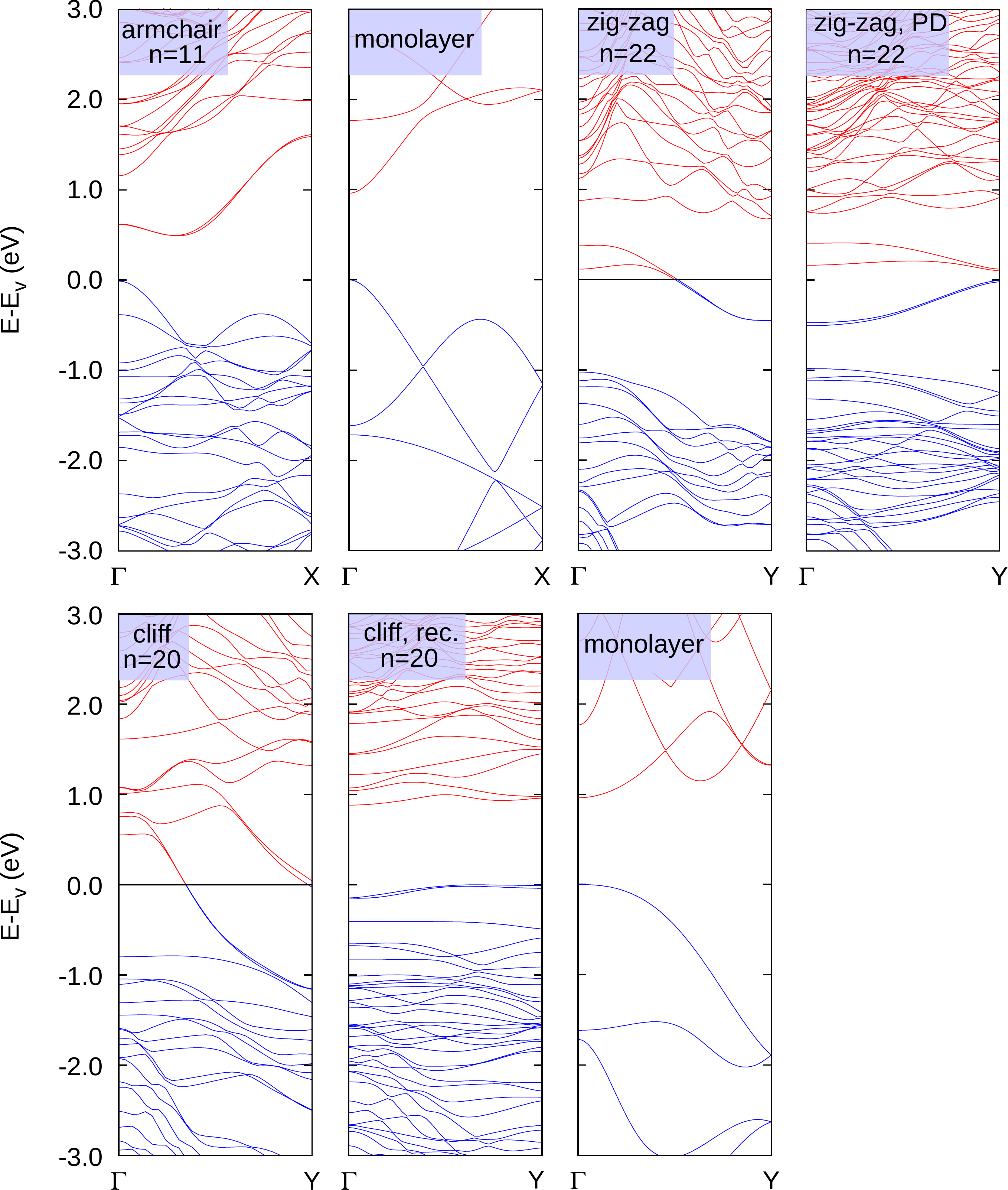}
\caption{(Color online)
Bandstructures of phosphorene nanoribbons.
The zero of the energy scale is set to the valence band top or to the Fermi level (for the metals).
On the right hand side, the bandstructure of
infinite monolayer phosphorene along the $\Gamma$-X and $\Gamma$-Y
directions is also shown for comparison.
}
\label{fig:bands}
\end{figure}

\begin{figure}
\includegraphics[width=8cm]{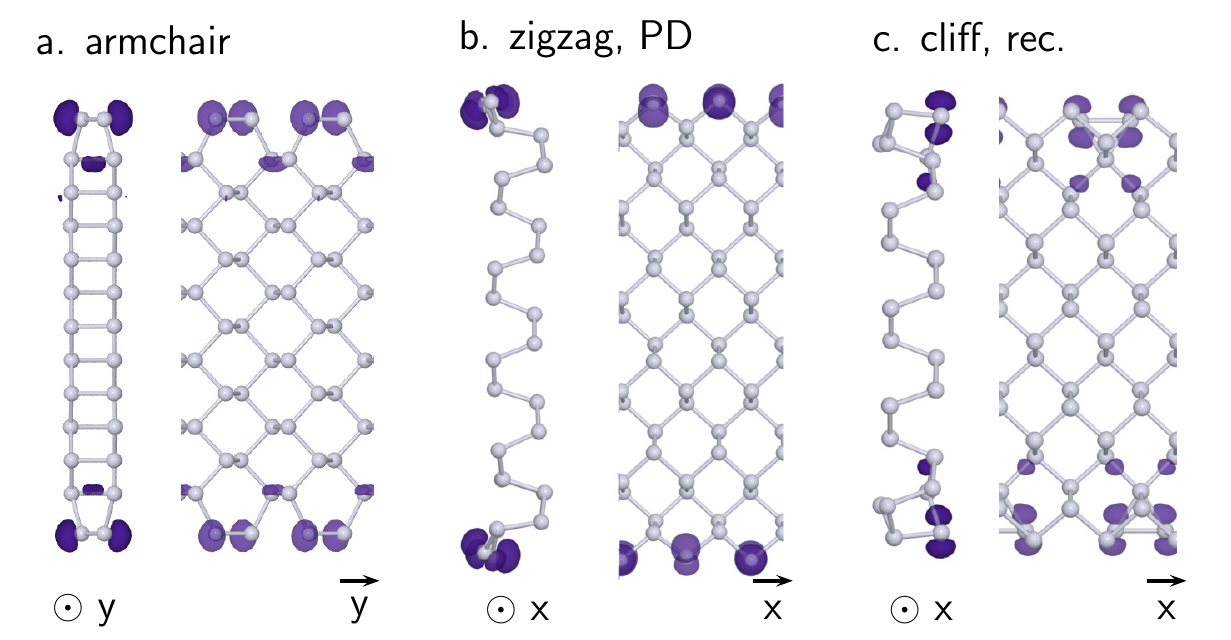}
\caption{(Color online)
Gap states of armchair ($n=13$), zigzag (Peierls-distorted, $n=22$) and cliff (reconstructed, $n=20$) nanoribbons with approximately the same width, 
highlighting the localization
on the edge atoms.
The plots are isosurface of the wavefunction square modulus for the lowest unoccupied state at $\Gamma$ ($n$=13) and for the highest occupied state at $\Gamma$ ($n=22$ and $n=20$).
\label{fig:cube}}
\end{figure}

\begin{figure}
\includegraphics[width=8cm]{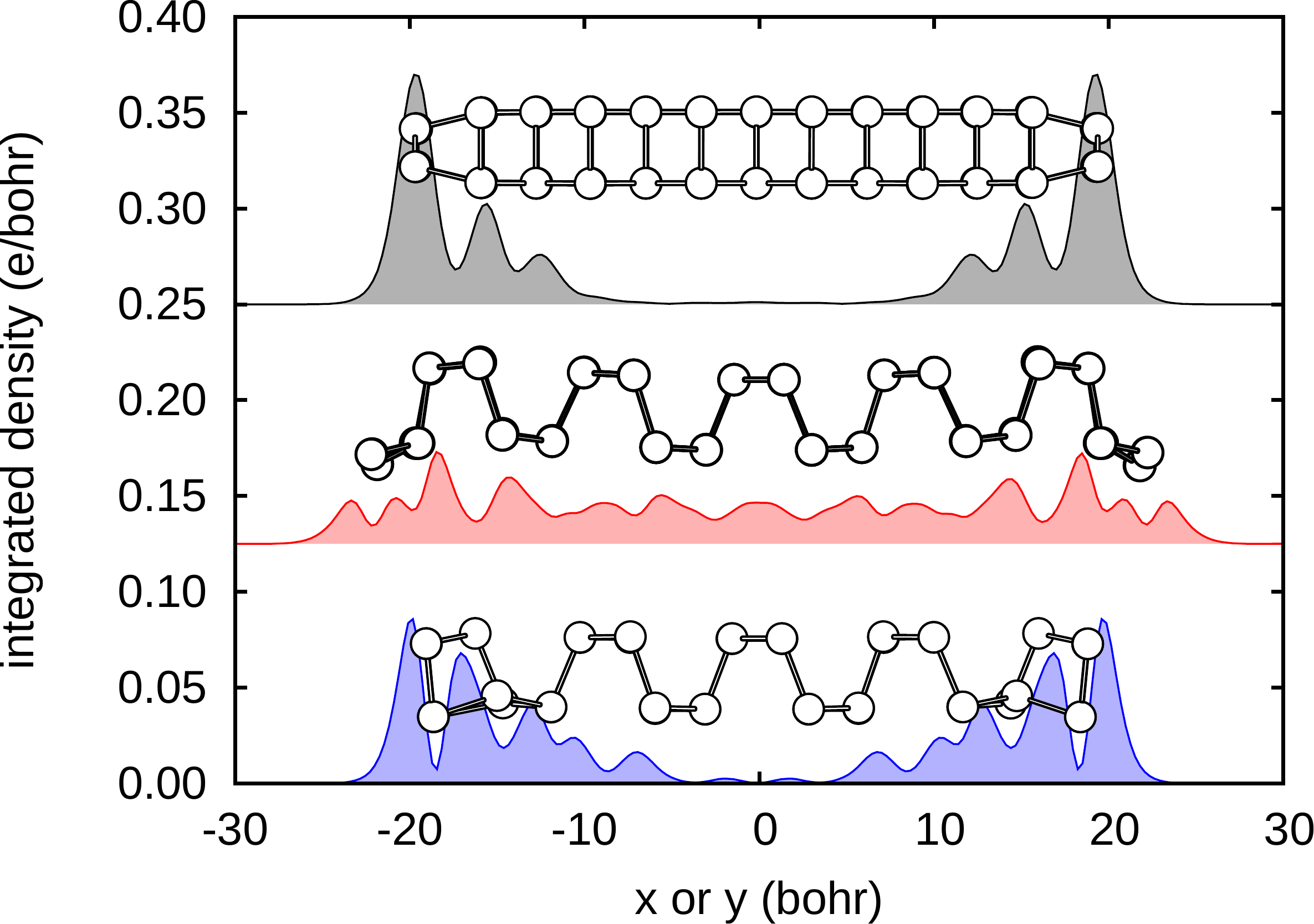}
\caption{(Color online)
Profile of the wavefunction square of gap states of armchair ($n=13$), 
zigzag (Peierls-distorted, $n=22$) and cliff (reconstructed, $n=20$) nanoribbons with approximately the same width, 
integrated over the perpendicular directions.
The states shown are the lowest unoccupied state at $\Gamma$ ($n=13$ and $n=22$) 
and for the highest occupied state at $\Gamma$ ($n=22$ and $n=20$).
\label{fig:density}}
\end{figure}

The presence of the edge boundary gives rise, in all three kinds of nanoribbons,
to gap states (Fig.~\ref{fig:bands}).
\footnote{See online supplemental material for bandstructures of armchair, zigzag and cliff nanoribbons 
as a function of size.}
The gap state of armchair nanoribbons is an edge state with
predominantly $p_z$ character (Fig.~\ref{fig:cube}), 
of which the amplitude decays very fast with the distance from the edge (Fig.~\ref{fig:density}).
The dispersion along the $x$ direction is approximately a concave-up parabola (except very close to $\Gamma$),
as expected from the two-band model.

In contrast, in unreconstructed zigzag and cliff-edge nanoribbons the Fermi level 
intersects the gap state, making them metallic.
However, this is changed by the structure reconstruction/distortion with unit cell doubling.

In the case of the zigzag phosphorene nanoribbon, 
the gap states are originally composed of $p_x$ and $p_z$ orbitals.
These nanoribbons have a pair of gap states, which are half-occupied.
The atoms in the origin of the gap states can be seen as
a linear system, and since they have half occupancy,
are sensitive to Peierls distortion with unit cell doubling.
The doubled states can be folded back into the smaller Brillouin Zone
and as a result of the distortion, a small gap is opened at the zone edge.
The opening of the gap between occupied and unoccupied states results in 
a small energy lowering ($<$ 40~meV).

The unoccupied gap state, similar to the original gap state
of the zigzag nanoribbon before Peierls distortion,
has higher amplitude at the edge, 
however it approaches that of a bulk state closer to the inner region (Fig.~\ref{fig:density}).
It can thus be considered a surface resonance,
or a softly confined state,\cite{borca-PRB-2010} rather than an edge state.
The occupied gap state, however, is an edge state (Fig.~\ref{fig:density}).

The cliff nanoribbon is insulating with a double degenerate edge-related state in the lower gap.
There is hybridization of the
edge states with the conduction band bottom close to $\Gamma$.
Thus, for some of the gap states the situation is intermediate between that
of a pure edge state and a surface resonance. 
The edge states separated in energy from 'crystal-like' states however are 
localized on the edge atoms (as the one shown in Fig.~\ref{fig:density}).

A difference between phosphorene nanoribbons and
their graphene counterpart is
that while zigzag graphene nanoribbons give rise to flat bands near the Fermi level
and as a result, the non-magnetic phase is unstable,
in phosphorene zigzag nanoribbons, 
the edge-related states have mixed $p_x$ and $p_z$ character.
They only become localized on the outer fridges 
for wide nanoribbons ($n>$20).
The respective bands
have dispersion in the $k_y$ direction (of the order of 1 eV to a few eV, depending on size).
Thus, the density of states is smaller.
Accordingly, we found no instability towards an anti-ferromagnetic phase.

\begin{figure}
\includegraphics[width=8cm]{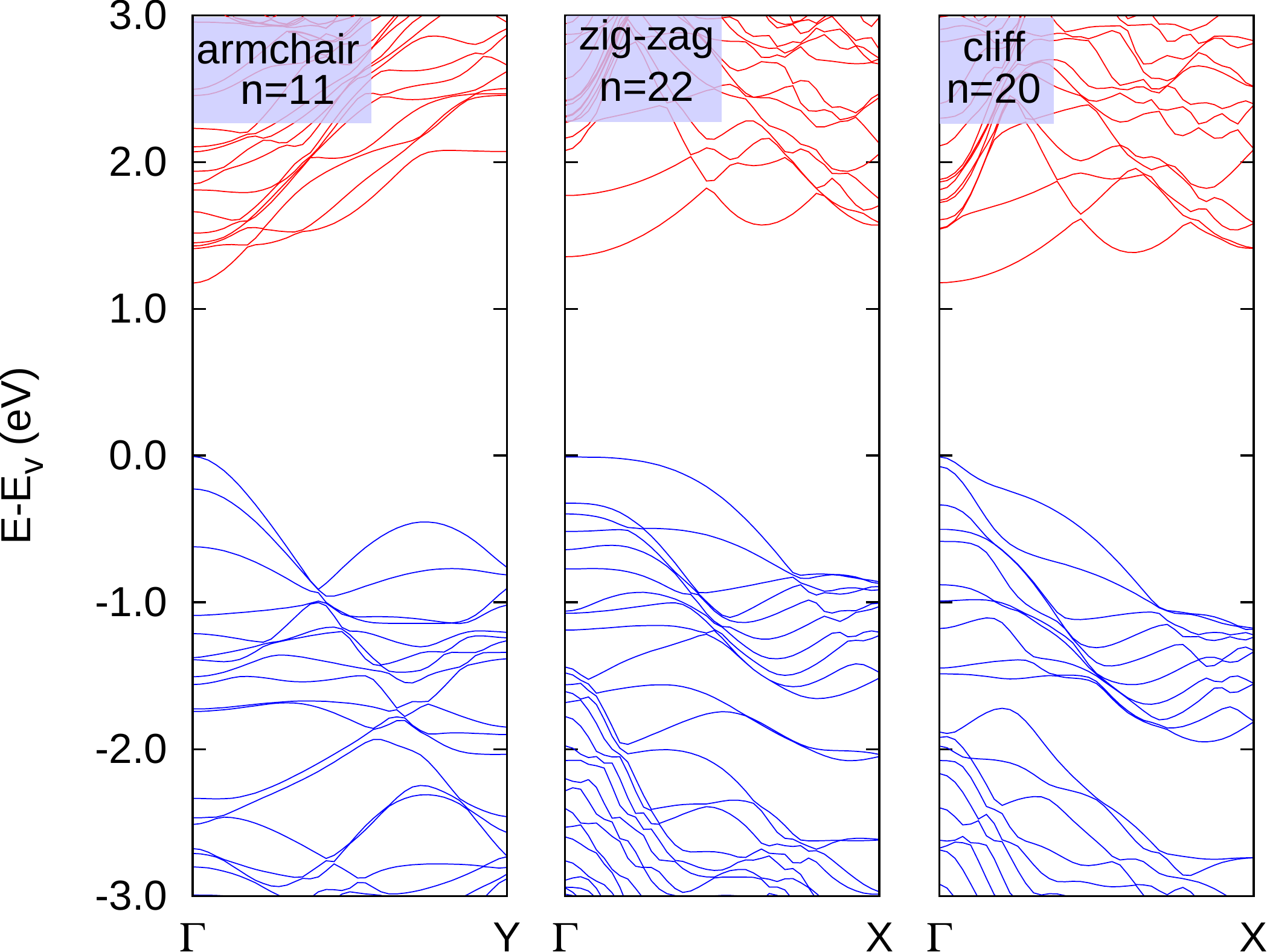}
\caption{(Color online)
Bandstructures of H-saturated phosphorene nanoribbons. 
The zero of the energy scale is set to the valence band top.
\label{fig:bands-H}}
\end{figure}

To confirm that the gap states originate on the edge, 
we examined the effect of saturated the edge atoms with hydrogen.
Each edge P was passivated with enough H to remove dangling bonds and restore three-fold coordination:
2 hydrogen atoms per edge P for armchair and cliff nanoribbons, and one hydrogen atom per edge P.
In all three kinds of nanoribbons, the passivation  
with hydrogen removes the gap states Fig.~\ref{fig:bands-H}.
As expected, the bandgap in this case decreases with increasing
system size.

Finally, in order to evaluate whether such edges can be found experimentally,
we look into their formation energy, defined as:
\begin{equation}
E_f=E_{\rm tot}-\sum_in_i\mu_i,
\end{equation}
where $E_{\rm tot}$ is the total energy of the unit cell,
and $n_i$ and $\mu_i$ are the number of atoms of the element $i$=$\{{\rm P,H}\}$ and
its chemical potential, respectively.
The chemical potentials for P and H are obtained assuming 
an environment rich in P$_4$ and PH$_3$.

\begin{figure}
\includegraphics[width = 8cm]{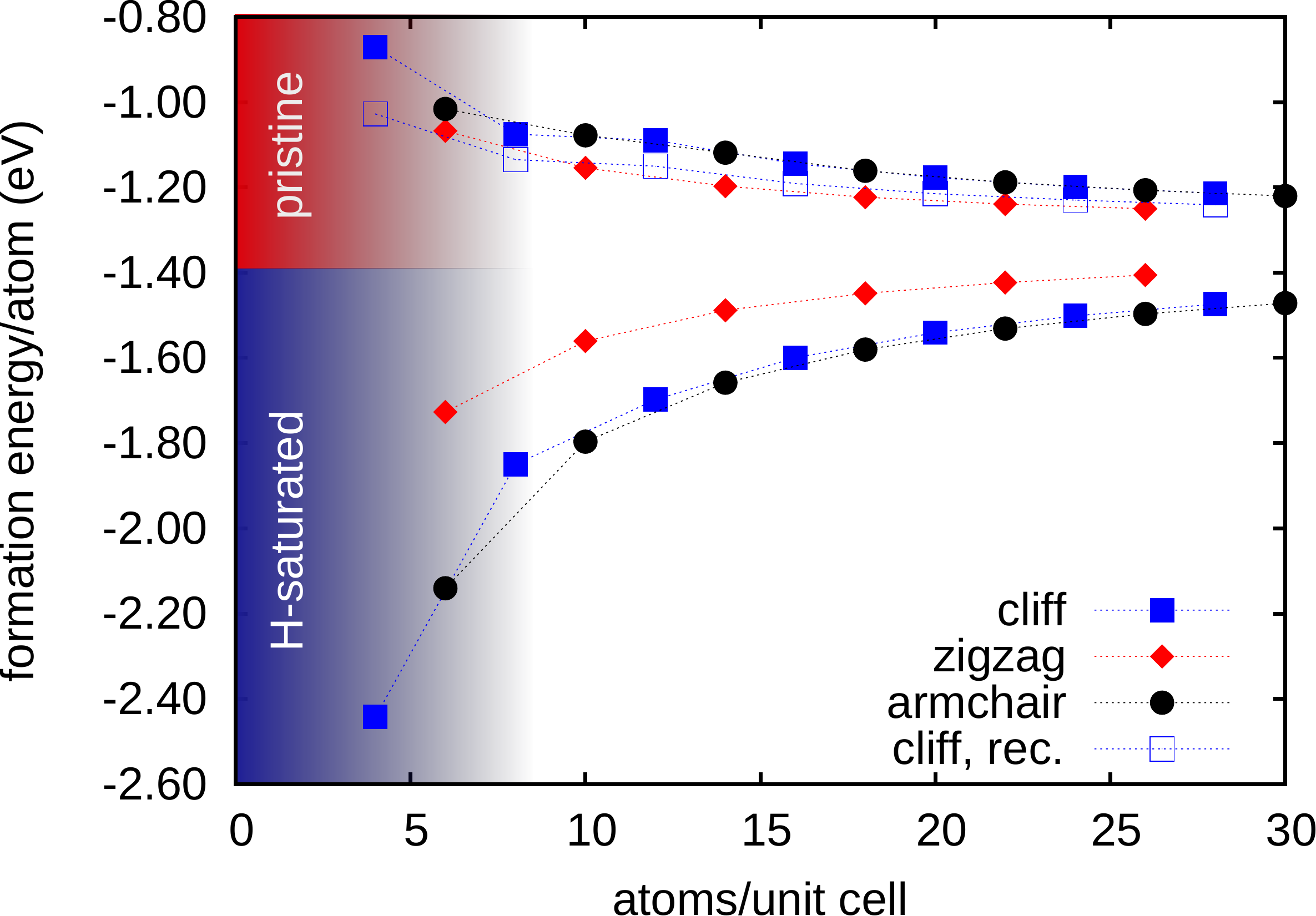}
\caption{(Color online)
Formation energy of the unsaturated (top) and saturated (bottom) nanoribbons,
in an environment rich in P$_4$  and PH$_3$, as a function 
of the number of phosphorus atoms.
For cliff-type nanoribbons the energies with a minimal unit cell
or with unit-cell doubling reconstruction (rec.) are both shown for comparison.
}
\label{fig:Ef}
\end{figure}

As black phosphorus crystals, phosphorene nanoribbons 
have a negative formation enthalpy with respect to white phosphorus.
In the unsaturated state, all three types of nanoribbons have
close formation energies, in a range of about 0.1 eV/atom (Fig.~\ref{fig:Ef}).
With hydrogen saturation, the formation energies follow a similar trend,
with the mono-saturated zigzag-edge nanoribbons hovering higher in energy by about 0.2~eV.
This energy difference might be the energy gain of forming a P-H bond 
from a P dangling orbital.

Figure ~\ref{fig:Ef}-a shows that the formation energy of the pristine nanoribbons increases, 
in absolute value, with the size, but the opposite is found for H-passivated nanoribbons.
This indicates that black-phosphorus monolayers will dissolve in reducing conditions.

In conclusion, phosphorene nanoribbons were found to be semiconducting, in their lowest energy structure,
with edge-induced states in the gap.
Armchair edges introduce edge states which decay exponentially with the distance to the edge.

In contrast, zigzag nanoribbons are metallic unless they undergo a Peierls distortion,
at low temperature, which opens a gap between occupied and unoccupied bands.
The highest occupied and lowest unoccupied bands originate on the edges
and can be described as a surface resonance and an edge state, respectively.
Since the distortion energy is very small ($<$40~meV), it is
likely that at high temperatures the edges present domain wall defects
or even become metallic.

Cliff-type edges also originate gap states, but those are less localized on the edge atoms
than the edge states of armchair nanoribbons.

An analytical description based on a NFEM provides further insight into
the origin of edge states, in particular in the case of armchair edges. 
Since this is simultaneously the type of edge less changed by reconstruction,
it is valid to generalise the DFT results obtained for narrow edges using this model.
In fact, within the NFEM it is possible to prove the existence of edge states for
non-interacting edges in armchair nanoribbons of any length.
This applies to a semi-infinite sheet,
which cannot be obtained from the DFT model.
The amplitude of these edge states was shown to decay exponentially toward the bulk,
in agreement with result of the first-principles calculations.

\acknowledgments            
A.S.R. acknowledges DOE grant DE-FG02-08ER46512, ONR grant MURI N00014-09-1-1063. A.H.C.N. acknowledges NRF-CRP award ``Novel 2D materials with tailored properties: beyond graphene" (R-144-000-295-281). The first-principles calculations were carried out on the GRC high-performance computing facilities.


\begin{thebibliography}{99}
\bibitem{footnote} 
The nanoribbons are modeled in a box.
The minimum edge-to-edge distance between neighboring images of the nanoribbons is 10\AA\ along the $z$ direction or 15\AA\ along the perpendicular direction.
Both the lattice parameter of the nanoribbon 
and the atomic coordinates are optimized to minimize the energy.
\end{thebibliography}
\end{document}